\documentclass[12pt]{article}

\usepackage[usenames,dvipsnames]{xcolor}
\usepackage{tikz}
\usepackage{pgfplots}
\pgfplotsset{compat=1.3}
\usepackage{subfigure}
\usepackage{array}
\newcolumntype{C}[1]{>{\centering\arraybackslash}p{#1}}

\usetikzlibrary{arrows,automata}
\usepackage{lscape}
\usepackage{float}
\usepackage[utf8]{inputenc}
\usepackage{authblk}
\usepackage[margin=1in]{geometry}
\usepackage{hyperref}
\usepackage{listings}
\usepackage{xcolor}
\usepackage{graphicx}
\usepackage{enumitem,amssymb}
\usepackage{amsmath}
\newlist{todolist}{itemize}{2}
\setlist[todolist]{label=$\square$}
\usepackage{pifont}
\definecolor{codegreen}{rgb}{0,0.6,0}
\definecolor{codegray}{rgb}{0.5,0.5,0.5}
\definecolor{codepurple}{rgb}{0.58,0,0.82}
\definecolor{backcolour}{rgb}{0.95,0.95,0.92}

\lstdefinestyle{mystyle}{
  backgroundcolor=\color{backcolour},   commentstyle=\color{codegreen},
  keywordstyle=\color{magenta},
  numberstyle=\tiny\color{codegray},
  stringstyle=\color{codepurple},
  basicstyle=\ttfamily\footnotesize,
  breakatwhitespace=false,         
  breaklines=true,                 
  captionpos=b,                    
  keepspaces=true,                 
  numbers=left,                    
  numbersep=5pt,                  
  showspaces=false,                
  showstringspaces=false,
  showtabs=false,                  
  tabsize=2
}

\providecommand{\keywords}[1]{\textbf{\textit{Keywords:}} #1}

\newcommand{\Rspace}{\textnormal{\sffamily R\space}}
\newcommand{\Cppspace}{\textnormal{\sffamily C++\space}}

\newcommand{\code}[1]{\texttt{#1}}

\usepackage{natbib}
\setcitestyle{authoryear,open={(},close={)}}

\usepackage{array,booktabs,csvsimple,longtable,tabularx,tabulary}

\makeatletter
\def\maxwidth{ %
	\ifdim\Gin@nat@width>\linewidth
	\linewidth
	\else
	\Gin@nat@width
	\fi
}
\makeatother

\definecolor{fgcolor}{rgb}{0.345, 0.345, 0.345}
\newcommand{\hlnum}[1]{\textcolor[rgb]{0.686,0.059,0.569}{#1}}%
\newcommand{\hlstr}[1]{\textcolor[rgb]{0.192,0.494,0.8}{#1}}%
\newcommand{\hlstd}[1]{\textcolor[rgb]{0.345,0.345,0.345}{#1}}%
\newcommand{\hlkwb}[1]{\textcolor[rgb]{0.69,0.353,0.396}{#1}}%
\newcommand{\hlkwc}[1]{\textcolor[rgb]{0.333,0.667,0.333}{#1}}%
\newcommand{\hlkwd}[1]{\textcolor[rgb]{0.737,0.353,0.396}{\textbf{#1}}}%

\usepackage{framed}
\makeatletter
\newenvironment{kframe}{%
	\def\at@end@of@kframe{}%
	\ifinner\ifhmode%
	\def\at@end@of@kframe{\end{minipage}}%
\begin{minipage}{\columnwidth}%
	\fi\fi%
	\def\FrameCommand##1{\hskip\@totalleftmargin \hskip-\fboxsep
		\colorbox{shadecolor}{##1}\hskip-\fboxsep
		\hskip-\linewidth \hskip-\@totalleftmargin \hskip\columnwidth}%
	\MakeFramed {\advance\hsize-\width
		\@totalleftmargin\z@ \linewidth\hsize
		\@setminipage}}%
{\par\unskip\endMakeFramed%
	\at@end@of@kframe}
\makeatother

\definecolor{shadecolor}{rgb}{.97, .97, .97}
\definecolor{messagecolor}{rgb}{0, 0, 0}
\definecolor{warningcolor}{rgb}{1, 0, 1}
\definecolor{errorcolor}{rgb}{1, 0, 0}
\newenvironment{knitrout}{}{} 

\usepackage{alltt}

\newcommand{\U}{\mathbf{U}}
\renewcommand{\H}{\mathbf{H}}

\newcommand{\G}{\mathbf{G}}

\newcommand{\given}{ \, | \,}

\renewcommand{\P}[1]{\mathbb{P}\left(#1\right)}

\title{\bf PanelPRO: A \Rspace package \\ for multi-syndrome, multi-gene risk modeling \\ for individuals with a family history of cancer}

\author[1]{Gavin Lee*}
\author[2]{Qing Zhang*}
\author[3,4]{Jane W. Liang}
\author[3,4]{Theodore Huang}
\author[1]{Christine Choirat}
\author[3,4]{Giovanni Parmigiani}
\author[3,4]{Danielle Braun}

\affil[1]{Swiss Data Science Center, ETH Z\"urich and EPFL, Lausanne, Switzerland}
\affil[2]{Broad Institute of MIT and Harvard, Cambridge, MA, USA}
\affil[3]{Department of Biostatistics, Harvard T.H. Chan School of Public Health, Boston, MA, USA}
\affil[4]{Department of Data Sciences, Dana-Farber Cancer Institute, Boston, MA, USA}
\affil[*]{Equal Contribution}

\date{\today}

\begin{document}

\maketitle

\newpage

\begin{abstract}

Identifying individuals who are at high risk of cancer due to inherited germline mutations is critical for effective implementation of personalized prevention strategies. Most existing models to identify these individuals focus on specific syndromes by including family and personal history for a small number of cancers. Recent evidence from multi-gene panel testing has shown that many syndromes once thought to be distinct are overlapping,  motivating the development of models that incorporate family history information on several cancers and predict mutations for more comprehensive panels of genes.

Once such class of models are Mendelian risk prediction models, which use family history information and Mendelian laws of inheritance to estimate the probability of carrying genetic mutations, as well as future risk of developing associated cancers. To flexibly model the complexity of many cancer-mutation associations, we present a new software tool called \textbf{PanelPRO}, an \Rspace package that extends the previously developed BayesMendel \Rspace package to user-selected lists of susceptibility genes and associated cancers. The model identifies individuals at an increased risk of carrying cancer susceptibility gene mutations and predicts future risk of developing hereditary cancers associated with those genes. Additional functionalities adjust for prophylactic interventions, known genetic testing results, and risk modifiers such as race and ancestry. The package comes with a customizable database with default parameter values estimated from published studies. This includes estimates of cancer penetrances, allele frequencies, and risk ratios for specific interventions, all stratified on different sets of demographic factors. 

The \textbf{PanelPRO} package is open-source and provides a fast and flexible back-end for multi-gene, multi-cancer risk modeling with pedigree data. 
The software enables the  identification of high-risk individuals, which will have an impact on personalized prevention strategies for cancer and individualized decision making about genetic testing. 

\end{abstract}


\clearpage

\section{Introduction}
\label{sec:introduction}

In the last decade, DNA sequencing has changed dramatically. Tests have become faster and more affordable, leading to discovery of a growing number of germline pathogenic variants associated with  increased cancer risk. Multi-gene panels are routinely available and include varying combinations of genes \citep{plinchta2016}. Evidence is accruing that gene mutations, which were typically believed to be only associated with one or two types of hereditary cancers, may in fact increase the risk for a wider range of syndromes. These advancements have changed the genetic counseling landscape by introducing a need to consider a wider set of individual genes and cancers to accurately assess overall risks. In the context of genetic counseling, the importance of accurate estimates of carrier probabilities is well-known \citep{nelson2014risk}. With the number of genes of interest and their combinations increasing, efficient calculation of these estimates becomes crucial in clinical settings. 

Existing Mendelian models consider a relatively narrow subset of cancers and genes. For example, BRCAPRO, available in the BayesMendel \Rspace package \citep{chen2004bayesmendel}, considers two cancers (breast and ovarian) and two genes (BRCA1 and BRCA2). Boadicea v4 Beta \citep{Boadicea:2020} considers BRCA1, BRCA2, PALB2, CHEK2 and ATM mutations in the same cancers. To comprehensively incorporate cancers, genes, and their interactions, we introduce \textbf{PanelPRO}, an \Rspace package which aims to efficiently and flexibly scale to the demands of germline panel testing. The newly developed package has  the following key advantages:
\begin{itemize}
    \item Customizable model specification, including the choice of genes and cancers included in the model;
    \item Customizable model parameters, including the allele frequencies and penetrances;
    \item Accurate default parameter estimates, curated from an extensive literature search;
    \item Flexibility to incorporate cancer risk modifiers such as prophylactic surgeries;
    \item Comprehensive user input checks for pedigrees;
    \item Speed of computation, through an optimized \Cppspace implementation of an efficient algorithm \citep{madsen2018efficient}
\end{itemize}

The newly developed \textbf{PanelPRO} model encompasses 25 genes and 18 cancers. These inputs are fully scalable (subject to reasonable run-times), and it is intended that these numbers become larger as more research on other genes becomes available. The package contains a comprehensive collection of functions designed to efficiently calculate carrier probabilities and future cancer risk for individuals, given detailed information about their family history. 

\textbf{PanelPRO} is fully back-compatible with the existing \textbf{BayesMendel} package; individual models within that package (for example, \textit{BRCAPRO}, \textit{MMRPRO}, \textit{BRCAPRO5} or \textit{BRCAPRO6}) can be called directly from \textbf{PanelPRO} by passing this model specification to the main function call. We expect that users of \textbf{BayesMendel} will migrate to this generalized and customizable enhancement.

\section{Background}
\label{sec:background}

In genetic counseling contexts, an individual may be suspected of inherited cancer susceptibility if their family history exhibits certain patterns. For example, if two or more relatives have the same type of cancer on the same side of the family, or if cancer diagnoses in the family are particularly early, they may be referred to testing for mutations in genes associated with increased risk for those specific cancers. For example, in the case of hereditary breast cancer, guidelines in the US were established to identify patients who have higher likelihoods of benefiting from germline genetic testing. Thresholds for testing were set high initially, since genetic testing was very expensive at the time \citep{manahan2019consensus}. Although cost of testing has decreased and guidelines are constantly changing, accurate calculation of carrier probabilities, given family history, is essential in supporting the decision for further testing, preventative treatment, or even family planning \citep{chen2004bayesmendel}.

\begin{table}[htbp]
    \centering
\begin{tabular}{lcp{3in}}
\toprule 
    Genotypes & & \\
       & $\G_i = (G_{ki})_{k=1}^K$ & genotype of individual $i$, where $G_{ki}$ is the binary indicator for carrying a deleterious mutation in the $k$th gene \\
        & $\G = (\G_i)_{i=1}^I$ & genotypes of all family members $i=1,\dots,I$ \\[10pt]
        \midrule
    Sex  & & \\
        & $U_i$ & binary indicator that individual $i$ is male \\
        & $\U = (U_i)_{i=1}^I$ & binary male indicators for all family members $i=1,\dots,I$ \\[10pt]
\midrule 
     Cancer history & & \\
& $T_{ri}$ & age of diagnosis of the $r$th cancer for individual $i$ \\
& $C_i$ & individual $i$'s censoring age (current age or age of death) \\
& $\delta_{ri} = I(T_{ri} \leq C_i)$ & binary indicator that cancer $r$ occurs before the censoring age for individual $i$ \\
& $\H_{ri} = (C_i, \delta_{ri}, T_{ri})$ & observed history of the $r$th cancer for individual $i$, not including risk modifiers and interventions \\
& $\H_i = (\H_{ri})_{r=1}^R$ & all observed history for individual $i$\\
& $\H = (\H_i)_{i=1}^I$ & observed histories for all family members $i=1,\dots,I$ \\ [4pt]
\cline{2-3} \\ [-8pt]
& $T_{dri}$ & individual $i$'s age of death from causes other than cancer $r$ \\
& $T_{ri}^* = \min(T_{ri}, T_{dri})$ & individual $i$'s age of first outcome, either cancer $r$ or death from causes other than cancer $r$ \\
 & $J_{ri} = I(T_{ri}^* = T_{ri})$ & binary indicator that individual $i$ develops the $r$th cancer \\
\bottomrule
\end{tabular}
    \caption{Notation for Mendelian Modeling for a model with $K$ genes and $R$ cancers and a family of $I$ members. The subscript $i$ denotes the $i$th family member. }
    \label{tab:notmedmod}
\end{table}

\subsection{Genotype probabilities}
\label{subsec:genotype}


 \textbf{PanelPRO} predicts an individual's probability of having a specified genotype. We use the notation in Table~\ref{tab:notmedmod}.
Without loss of generality, let the subscript $i=1$ represent  the counselee (i.e. the individual who is counseled). For simplicity, we only consider one counselee, though the model can handle multiple counselees in a computationally efficient manner. The counselee's genotype probability is
\begin{equation}
	\label{eq:genotype_distribution}
	\P{\G_1 \given \H, \U}.
\end{equation}
Using Bayes' rule, the law of total probability and the assumption of independence of family phenotypes given genotypes and sex, this can be written as
\begin{align}
    \label{eq:genotype_distribution_rewritten}
    \nonumber \P{\G_1 \given \H, \U} &\propto \P{\G_1} \sum_{\G_2, \dots, \G_I} \prod_{i=1}^I \P{\H_i \given \G_i, U_i} \P{\G_2, \dots, \G_I \given \G_1} \\
    &= \P{\G_1} \sum_{\G_2, \dots, \G_I} \prod_{r=1}^R \prod_{i=1}^I \P{\H_{ri} \given \G_i, U_i} \P{\G_2, \dots, \G_I \given \G_1}.
\end{align}
\sloppy From this representation of the posterior probability, we can clearly see the model and user inputs to \textbf{PanelPRO}. $\P{\G_1}$ represents the allele frequencies for each gene in the model. $\P{\H_{ri} \given \G_i, U_i}$ are derived from the cancer penetrances $\P{T_{ri} = t \given \G_i, U_i}$. Explicitly,
\[
\P{\H_{ri} \given \G_i, U_i} =
\begin{cases}
1 - \sum_{s=1}^{C_i} \P{T_{ri} = s \given \G_i, U_i} & \text{ if } \delta_{ri} = 0 \\
\P{T_{ri} = T_{ri}^{obs} \given \G_i, U_i} & \text{ if } \delta_{ri} = 1
\end{cases}
\]
where $T_{ri}$ is the random variable and $T_{ri}^{obs}$ is the observed cancer age. By default, the allele frequencies and penetrances are obtained from existing peer-reviewed studies and estimates, but are completely customizable within \textbf{PanelPRO}.

Since the genotype space $\{(\G_2, \dots, \G_I) : \G_i \in \{0, 1\}^K, i = 2, \dots, I\}$ is large for large values of $K$, we use the peeling-paring algorithm \citep{madsen2018efficient} as an approximation, only allowing a pre-specified number of mutations to be simultaneously present in the same individual. The pedigree structure from the user input is used to derive the $\P{\G_2, \dots, \G_I \given \G_1}$ term in Equation \ref{eq:genotype_distribution_rewritten} using Mendelian laws of inheritance.

\subsection{Future cancer risk}
\label{subsection:futurerisk}

 \textbf{PanelPRO} also estimates  future cancer risk, based on the previously calculated genotype distribution of the individual.
Suppose the counselee has not developed the $r$th cancer by the current age. Then the risk of developing the $r$th cancer in $t_0$ years is
\begin{align}
    \label{eq:crude_future_risk}
    \P{T_{r1}^* \leq C_i + t_0, J_{ri} = 1 \given \H, \U} = \sum_{\G_1} \P{T_{r1}^* \leq C_i + t_0, J_{ri} = 1 \given \G_1, \U} \P{\G_1 \given \H, \U}.
\end{align}

Equation~(\ref{eq:crude_future_risk}) produces  so-called `crude' risk, since competing risks of death from causes other than the specified cancer are accounted for. Thus, the reported future risk is the probability that the counselee develops the $r$th cancer within the next $t_0$ years and does not die from other causes beforehand, given the cancer history and sexes of the family. $\P{T_{r1}^* \leq C_i + t_0, J_{ri} = 1 \given \G_1, \U}$ is the crude penetrance and is also a model input with default values estimated from the literature.

\textbf{PanelPRO} also provides the option to report `net' future risk, which is the probability that the counselee develops the $r$th cancer in a hypothetical world where they cannot die from other causes, given the cancer history and sexes of the family. This risk type is not as realistic but some clinicians find it useful, as it focuses on the specified cancer and allows them to factor qualitatively the patient-specific covariates that may affect the patient's risk. To report net future risk, \textbf{PanelPRO} uses the net penetrances $\P{T_{ri} = t \given \G_i, U_i}$. Note that the genotype probabilities in expression~(\ref{eq:genotype_distribution_rewritten}) were calculated using net penetrances, as we do not collect death from other causes as a user input.

\section{Package Workflow}
\label{sec:workflow}

\begin{figure}[h!]
    \centering
    \includegraphics[width=.95\textwidth]{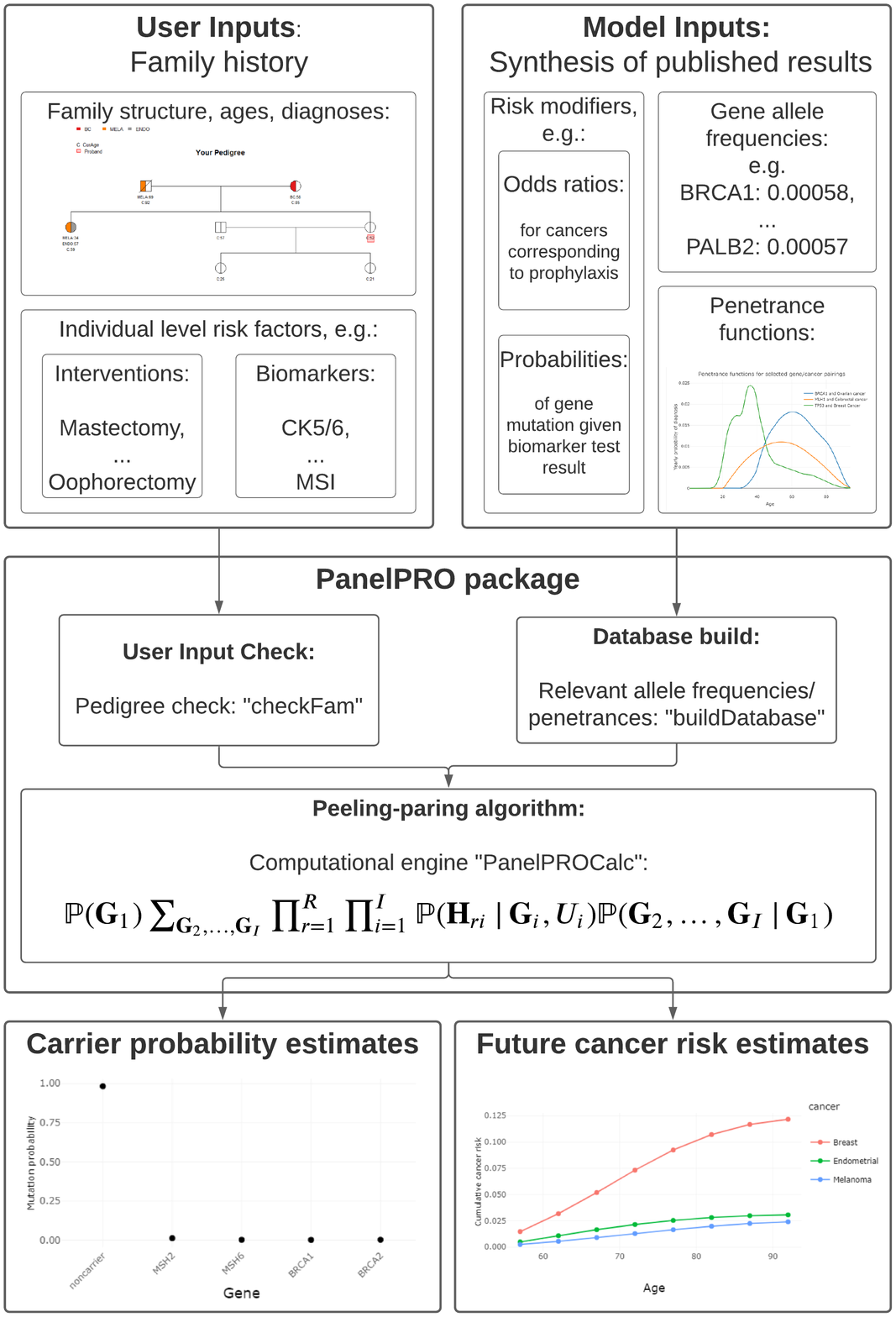}
    \caption{\textbf{PanelPRO} Package workflow.}
    \label{fig:package_workflow}
\end{figure}

The workflow of the package includes four main parts: the input including user and model input;  pre-processing of the inputs including user input checks and a database build; running the peeling-paring algorithm; and outputting the results. Figure \ref{fig:package_workflow} shows the workflow.

\subsection{Input}
\label{subsec:input}

\subsubsection*{User Input}
\label{subsubsec:user_input}

\begin{figure}[h!]
    \centering
    \includegraphics[width=.9\textwidth]{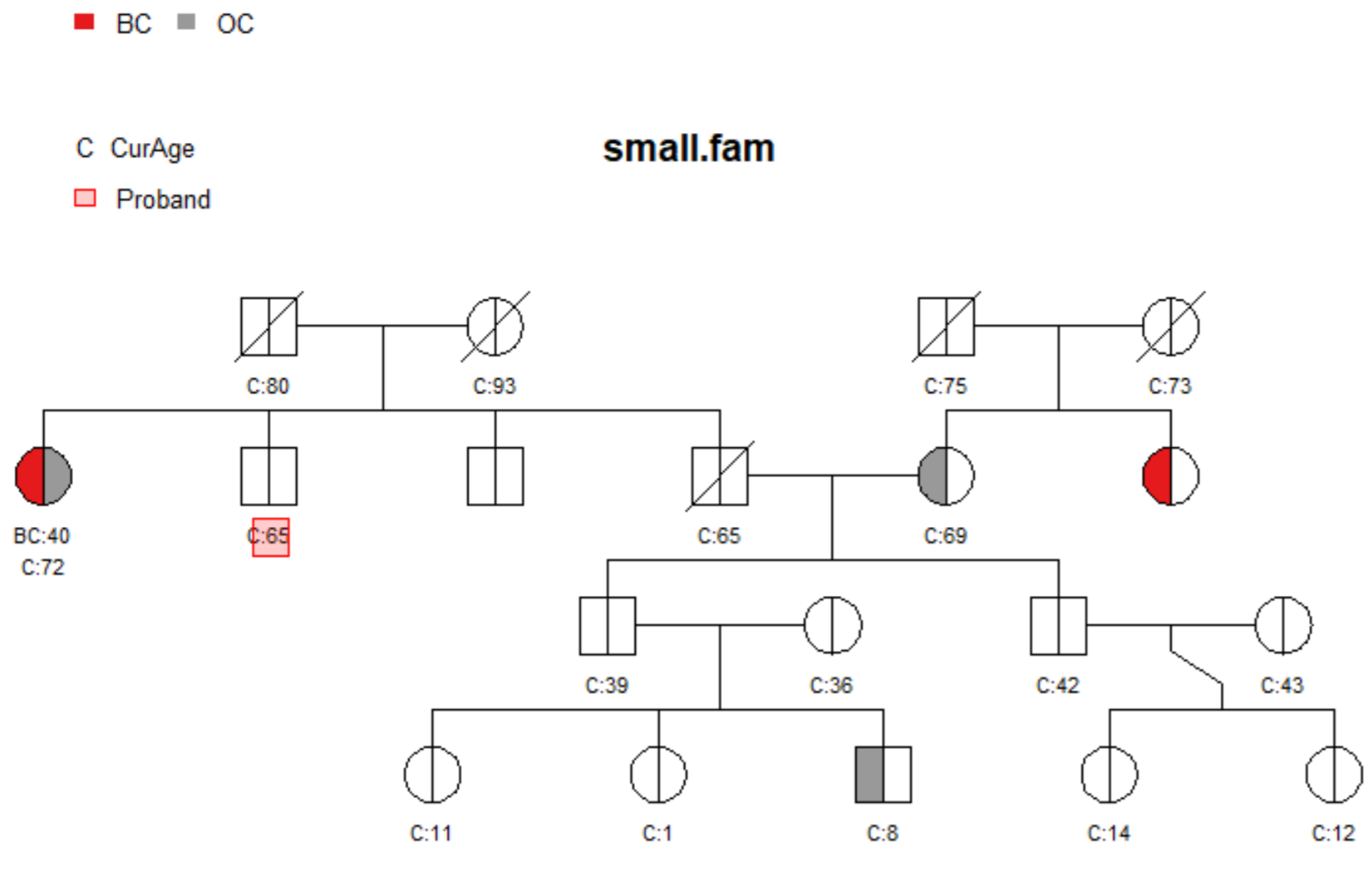}
    \caption{\code{small.fam} sample pedigree as included in the \textbf{PanelPRO} package, plotted using the external \code{visPed} package. The colors refer to cancer diagnoses in the legend. The age of diagnosis is shown below the individual if it is known.}
    \label{fig:small.fam}
\end{figure}

The main input from the user is their pedigree. This is in the form of a \Rspace \code{data.frame} which contains detailed information about known family members, such as their ages and previous cancer diagnoses. The pedigree structure is defined by the \code{ID}, \code{MotherID} and \code{FatherID} columns. Previous cancer diagnoses and their ages of diagnosis are stored in the \code{isAff*} and \code{Age*} columns, respectively, where \code{*} represents a cancer type, designated according to a standard nomenclature of two- to four- letter tags which are also used in visualizations. Risk modifiers, such as prophylactic surgeries, can be incorporated to adjust the likelihood calculation. Previous genetic testing history can also be incorporated. The package contains three sample pedigrees, called \code{small.fam}, \code{fam10}, and \code{fam25}, which illustrate most of the data that can be included in the user input. A clipped version (with a subset of the necessary columns) of \code{small.fam} pedigree is shown below.
\newpage
\begin{knitrout}
\definecolor{shadecolor}{rgb}{0.969, 0.969, 0.969}\color{fgcolor}\begin{kframe}
\begin{alltt}
\hlkwd{str}\hlstd{(small.fam)}
\end{alltt}
\begin{alltt}
\hlstd{small.fam}
\end{alltt}
\small
\begin{verbatim}
##   ID Sex MotherID FatherID isProband CurAge isAffBC AgeBC      riskmod BRCA1
## 1  1   0       NA       NA         0     93       0    NA                 NA
## 2  2   1       NA       NA         0     80       0    NA                  1
## 3  3   0        1        2         0     72       1    40   Mastectomy    NA
## 4  4   1        1        2         0     65       0    NA                  0
## 5  5   1        1        2         1     65       0    NA                 NA
## 6  6   1        1        2         0     NA       0    NA Hysterectomy    NA
\end{verbatim}
\end{kframe}
\end{knitrout}
The full specification of the pedigree structure is shown in Table \ref{tab:pedigree_structure}. The family tree pedigree can also be visualized by using the \textbf{\code{visPed}} package as in Figure~\ref{fig:small.fam}. This external package is available through \url{https://github.com/gavin-k-lee/visPed} and is based on the \code{kinship2} package available in CRAN \citep{sinnwell2014kinship2}.  \textbf{PanelPRO} itself does not contain pedigree plotting functionality. However, users can easily acquire the \textbf{\code{visPed}} package separately.

\begin{table}[h!]
	\centering
\begin{tabular}{p{0.2\textwidth}p{0.45\textwidth}p{0.3\textwidth}}
	\toprule
	Column & Definition & Value \\ 
	\midrule 
	\code{ID} & unique numeric identifier of each individual & non-repeated strictly positive integer \\
	
	\code{MotherID} & \code{ID} of one's mother & strictly positive integer or \code{NA} (missing) \\
	
	\code{FatherID} & \code{ID} of one's father & strictly positive integer or \code{NA} (missing) \\
	
	\code{Sex} & sex of the individual: 1 for male, 0 for female & one of $\{0, 1, \code{NA} \}$ \\
	
	\code{isProband} & indicates the proband or counselee by 1 and 0 otherwise -- multiple probands can be specified & one of $\{0, 1\}$ \\
	
	\code{CurAge} & age of censoring: either the current age or death age, depending on \code{isDead} status & positive integer or \code{NA} (missing) \\
	
	\code{isAff*} & affection status of cancer \code{*} & one of $\{0, 1\}$ \\
	
	\code{Age*} & affection age of cancer \code{*}. & positive integer or \code{NA} (missing) \\
	
	\code{isDead} & whether someone has died & one of $\{0, 1, \code{NA}\}$ \\
	
	\code{race} & race of individual (used to modify penetrance) & one of \code{All\_Races}, \code{AIAN}, \code{Asian}, \code{Black}, \code{White}, \code{Hispanic}, \code{WH}, \code{WNH}, \code{NA} \\
	
	\code{Ancestry} & ancestry of individual (used to modify allele frequencies) & one of \code{AJ}, \code{nonAJ}, \code{Italian}, \code{NA} \\
	
	\code{Twins} & identifies siblings who are identical twins or multiple births & each set is identified by a unique integer, and 0 otherwise \\
	
	\code{riskmod} & preventative interventions which modify penetrance & \code{list}, combination of  \code{Mastectomy}, \code{Hysterectomy}, \code{Oophorectomy} \\
	
	\code{InterAge} & age of each preventative interventions & \code{list}, combination of integers \\
	
	Gene name from \code{GENE\_TYPES} & germline testing result & one of $\{0, 1, \code{NA} \}$ \\
	
	Marker name from \code{CK14}, \code{CK5.6}, \code{ER}, \code{PR}, \code{HER2}, \code{MSI} & marker testing result & one of $\{0, 1, \code{NA} \}$ \\
	
	\bottomrule
\end{tabular}
	\caption{Pedigree structure in \textbf{PanelPRO}.}
	\label{tab:pedigree_structure}
\end{table}

The user defines the cancers in the pedigree, as well as the genes for which carrier probabilities should be returned, in the main \code{PanelPRO} function call.
\begin{knitrout}
\definecolor{shadecolor}{rgb}{0.969, 0.969, 0.969}\color{fgcolor}\begin{kframe}
\begin{alltt}
\small
\hlkwd{PanelPRO}\hlstd{(}\hlkwc{pedigree} \hlstd{= small.fam,}
         \hlkwc{cancers} \hlstd{=} \hlkwd{c}\hlstd{(}\hlstr{"Breast"}\hlstd{,} \hlstr{"Ovarian"}\hlstd{),}
         \hlkwc{genes} \hlstd{=} \hlkwd{c}\hlstd{(}\hlstr{"BRCA1"}\hlstd{,} \hlstr{"BRCA2"}\hlstd{,} \hlstr{"ATM"}\hlstd{,} \hlstr{"MSH2"}\hlstd{))}
\end{alltt}
\end{kframe}
\end{knitrout}

The user can also select other options in the function call. Examples include the maximum number of simultaneous gene mutations considered for a given individual, whether a parallelized version of the algorithm is performed, and the number of imputations in case of missing data (see section \ref{subsec:imputation}). Many of the useful options are listed in Table~\ref{tab:options}.

\begin{table}[h!]
\begin{tabularx}{\linewidth}{p{2.5cm}p{2.5cm}p{3.5cm}X}
\toprule
Option & Default Value & Possible values & Description \\
\midrule 

\code{max.mut} & \code{NULL} & Integers up to the number of genes & Number of maximum simultaneous mutations, also known as the paring parameter. If no integer has been input, it re-defaults to 2.\\

\code{iterations} & 20 & Integers from 1 upwards & In case of missing current or cancer ages in the pedigree, this is the number of times those ages will be imputed. \\

\code{parallel} & \code{TRUE} & \code{TRUE} or \code{FALSE} & If age imputations are needed, this parameter can be set to utilize multiple cores on one's machine. \\

\code{net} & \code{FALSE} & \code{TRUE} or \code{FALSE} & Determines whether net or crude penetrances are used to compute future risk of cancer. Net penetrances exclude all other causes of death, apart from the affected cancer. \\

\code{age.by} & 5 & Integers from 1 upwards & The intervals of age used to report the future risk of cancer. \\

\bottomrule
\end{tabularx}
\vspace{0.1cm}
\caption{List of options that the user can pass to \textbf{PanelPRO}, along with their defaults.}
\label{tab:options}
\end{table}

Passing these options to the function call is simple, as shown below.

\begin{knitrout}
\definecolor{shadecolor}{rgb}{0.969, 0.969, 0.969}\color{fgcolor}\begin{kframe}
\begin{alltt}
\hlkwd{PanelPRO}\hlstd{(}\hlkwc{pedigree} \hlstd{= small.fam,}
         \hlkwc{cancers} \hlstd{=} \hlkwd{c}\hlstd{(}\hlstr{"Breast"}\hlstd{,} \hlstr{"Ovarian"}\hlstd{),}
         \hlkwc{genes} \hlstd{=} \hlkwd{c}\hlstd{(}\hlstr{"BRCA1"}\hlstd{,} \hlstr{"BRCA2"}\hlstd{,} \hlstr{"ATM"}\hlstd{,} \hlstr{"MSH2"}\hlstd{),}
         \hlkwc{max.mut} \hlstd{=} \hlnum{1}\hlstd{,}
         \hlkwc{parallel} \hlstd{=} \hlnum{FALSE}\hlstd{)}
\end{alltt}
\end{kframe}
\end{knitrout}

Prophylactic surgeries (mastectomy, oophorectomy and hysterectomy) act as risk modifiers. They are specified in a column of  lists in the user input pedigree. Including these risk modifiers changes the resulting carrier probability and future risk outputs (see \ref{subsec:output}). Previous history of biomarker testing for breast and colorectal cancers can also be included in the model.

\subsubsection*{Model Input}
\label{subsubsec:model_input}

Calculation of carrier probabilities requires information about allele frequencies and penetrances for the mutations and cancers requested in the function call. They are derived from peer-reviewed studies whose results are catalogued in the \code{PanelPRODatabase}. At each function call, the code extracts the appropriate subset of gene-cancer combinations. Users are also free to change the defaults for their own purposes. The structure of this database is an \Rspace list. A partial output is provided below.

\begin{knitrout}\footnotesize
\definecolor{shadecolor}{rgb}{0.969, 0.969, 0.969}\color{fgcolor}\begin{kframe}
\begin{alltt}
\hlkwd{str}\hlstd{(PanelPRODatabase)}
\end{alltt}
\begin{verbatim}
## List of 7
##  $ Penetrance      : num [1:18, 1:26, 1:8, 1:2, 1:94, 1:2] 3.98e-05 2.80e-07 0.00 ...
##   ..- attr(*, "dimnames")=List of 6
##   .. ..$ Cancer   : chr [1:18] "Brain" "Breast" "Cervical" "Colorectal" ...
##   .. ..$ Gene     : chr [1:26] "APC" "ATM" "BARD1" "BMPR1A" ...
##   .. ..$ Race     : chr [1:8] "All_Races" "AIAN" "Asian" "Black" ...
##   .. ..$ Sex      : chr [1:2] "Female" "Male"
##   .. ..$ Age      : chr [1:94] "1" "2" "3" "4" ...
##   .. ..$ PenetType: chr [1:2] "Net" "Crude"
##  $ AlleleFrequency : num [1:25, 1:3] 1.45e-04 1.90e-03 3.41e-04 2.17e-05 1.37e-02 ...
##   ..- attr(*, "dimnames")=List of 2
##   .. ..$ Gene    : chr [1:25] "APC" "ATM" "BARD1" "BMPR1A" ...
##   .. ..$ Ancestry: chr [1:3] "AJ" "nonAJ" "Italian"
\end{verbatim}
\end{kframe}
\end{knitrout}

\subsection{Preprocessing}
\label{subsec:preprocessing}

\subsubsection*{Pedigree check}
\label{subsubsec:pedigree_check}

\begin{figure}[h!]
    \centering
    \includegraphics[width=.8\textwidth]{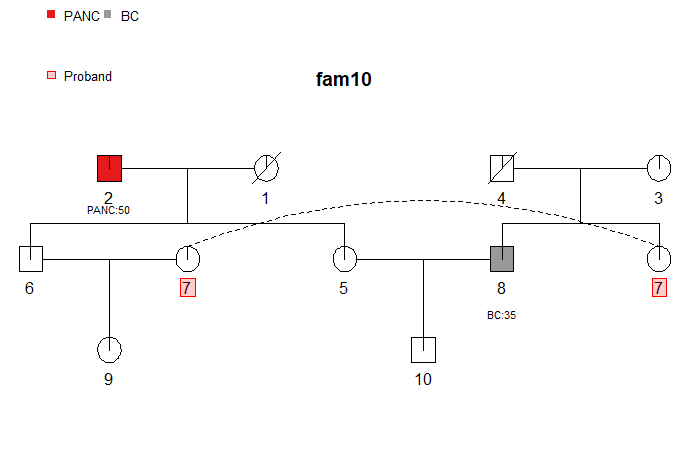}
    \caption{The sample pedigree \code{fam10} which contains a pedigree loop, due to the mating pattern of \code{6} and \code{8}, and \code{5} and \code{7}.}
    \label{fig:fam10}
\end{figure}

First, \textbf{PanelPRO} checks the structure of the user-supplied \Rspace \code{data.frame} containing the family history to be evaluated. Warnings are given to the user if certain values have been automatically changed to allow for accurate calculations. Errors are given if inconsistencies or ambiguities in the pedigree cannot be resolved such that the pedigree can be safely passed into downstream functions. Most of the checks are for consistency in terms of cancers and sex, as well as race or ancestry of parents/children and twins. The pedigree is also checked for `loops' which \textbf{PanelPRO} currently does not support. For example, within \code{fam10}, there are two sets of male and female siblings (four individuals) who have mated with the corresponding siblings in another family, as in Figure~\ref{fig:fam10}. This mating configuration results in a loop. For a more detailed definition of loops, see \cite{Fernando1993}. In addition, disconnected family members are detected and removed from the pedigree if they will not influence the counselee's results.

\subsubsection*{Build database}
\label{subsubsec:build_database}

Depending on the configuration of the model requested (cancers in the family, genes considered), a subset of \code{PanelPRODatabase} or a user-modified database will be created and passed through for further calculation. 

\subsection{Algorithm}
\label{subsec:algorithm}

The checked pedigree and \code{PanelPRODatabase} subset, as well as any user options, are then passed to the `peeling-paring' algorithm, which approximates Equation \ref{eq:genotype_distribution_rewritten}. It is based on the `peeling' algorithm as introduced by \cite{elston1971general} with its implementation based on \cite{Fernando1993}. The `paring' aspect of the algorithm limits the number of simultaneous mutations allowed. This is called the paring parameter and has a default value of 2. This results in an approximation which has been shown to be adequate for clinical purposes \citep{madsen2018efficient}. When the paring parameter is set equal to the number of distinct genes to be considered, the calculation is exact (assuming no other missing information about the pedigree). Future cancer risks are then calculated based on the law of total probability, using the previously calculated posterior carrier probabilities, as described in Section~\ref{sec:background}. These two calculations are performed in \code{PanelPROCalc}, as listed in Table \ref{tab:PanelPRO_functions}. The underlying algorithm is written in \textbf{Rcpp} using the \textbf{RcppArmadillo} package \citep{RcppArmadillo, Rcpp}. It uses, as much as possible, optimized data structures, vectorized operations and in-place modifications to be both time and memory efficient. See Figure~\ref{fig:run_times} for benchmarks on the run-time of the implementation.

The recursive nature of the peeling-paring algorithm allows for multiple counselee's to be specified in the function call without significant increase in the computational time. This is an advantage when multiple family members are at high risk and would benefit from knowing their carrier probabilities and future cancer risks.

\subsubsection*{Missing Data}
\label{subsec:imputation}

\code{PanelPRO} calculates mutation carrier probabilities for one or more  counselees. 
The peeling-paring algorithm requires both parents of the counselee to be present in the pedigree in order to link individuals who are non-founders. When there is only data for a single parent (whose children influence the results), we add a pseudo-parent who has the same prior allele frequencies as the parent for whom we do have information on. This allows the peeling-paring algorithm to run and serves as an approximation of the final results.

When the current age or age of cancer diagnosis of a family member is unknown, we use a multiple imputation procedure to repeatedly sample their age \citep{biswas2013simplifying}. Unknown current ages are sampled based on the current ages of the relatives, and unknown ages of cancer diagnosis are sampled from the penetrance, using the current age as an upper bound. The optional \code{impute.times} argument in \code{PanelPRO()} can be used to set the number of samples taken. The value labeled `estimate' in the output is the average of the results over the sampled ages, whilst the `lower' and `upper' bounds are the minimum and maximum values over the respective samples (whether it be for the posterior probabilities or future risks).

When \code{impute.times} is high (say, 50 or more), it is recommended to set the parameter \code{parallel} to TRUE. The algorithm will then use the \code{foreach} package and the existing cores in one's machine to execute the imputations in a parallel fashion, instead of sequentially, thereby speeding up the computation.

\subsection{Output}
\label{subsec:output}

For each proband in the pedigree, the output consists of
\begin{itemize}
    \item estimates of carrier probabilities,
    \item lower and upper bound estimates of carrier probabilities if imputations were made for missing data,
    \item estimates of future risks of cancers in 5-year intervals (the user can also change the length of the intervals), 
    \item lower and upper bound estimates of future risks of cancers in 5-year intervals if imputations were made for missing data.
\end{itemize}

Warning messages generated from \code{checkFam} have been omitted in the example below for brevity.

\begin{knitrout}
\definecolor{shadecolor}{rgb}{0.969, 0.969, 0.969}\color{fgcolor}\begin{kframe}
\begin{alltt}
\hlstd{output} \hlkwb{<-} \hlkwd{PanelPRO}\hlstd{(}\hlkwc{pedigree} \hlstd{= small.fam,}
                   \hlkwc{cancers} \hlstd{=} \hlkwd{c}\hlstd{(}\hlstr{"Breast"}\hlstd{,} \hlstr{"Ovarian"}\hlstd{),}
                   \hlkwc{genes} \hlstd{=} \hlkwd{c}\hlstd{(}\hlstr{"BRCA1"}\hlstd{,} \hlstr{"BRCA2"}\hlstd{,} \hlstr{"ATM"}\hlstd{,} \hlstr{"MSH2"}\hlstd{),}
                   \hlkwc{max.mut} \hlstd{=} \hlnum{2}\hlstd{,}
                   \hlkwc{parallel} \hlstd{=} \hlnum{FALSE}\hlstd{)}
\end{alltt}

{\ttfamily\noindent\itshape\color{messagecolor}{\#\# Your model has 2 cancers - Breast,Ovarian and 4 genes - BRCA1,BRCA2,ATM,MSH2}}

\begin{alltt}
\hlstd{output}
\end{alltt}
\begin{verbatim}
## $posterior.prob
## $posterior.prob$`5`
##          genes     estimate        lower        upper
## 1   noncarrier 0.0000000000 0.0000000000 0.0000000000
## 2        BRCA1 0.9959842769 0.9933338937 0.9962821073
## 3        BRCA2 0.0000000000 0.0000000000 0.0000000000
## 4          ATM 0.0000000000 0.0000000000 0.0000000000
## 5         MSH2 0.0000000000 0.0000000000 0.0000000000
## 6  BRCA1.BRCA2 0.0001276479 0.0001158473 0.0001881287
## 7    BRCA1.ATM 0.0025962774 0.0025557658 0.0026018069
## 8    BRCA2.ATM 0.0000000000 0.0000000000 0.0000000000
## 9   BRCA1.MSH2 0.0012917977 0.0009957958 0.0039622047
## 10  BRCA2.MSH2 0.0000000000 0.0000000000 0.0000000000
## 11    ATM.MSH2 0.0000000000 0.0000000000 0.0000000000
## 
## 
## $future.risk
## $future.risk$`5`
## $future.risk$`5`$Breast
##   ByAge    estimate       lower       upper
## 1    70 0.003093084 0.003092784 0.003094621
## 2    75 0.006725350 0.006724707 0.006728641
## 3    80 0.010882469 0.010881445 0.010887720
## 4    85 0.015564430 0.015562985 0.015571837
## 5    90 0.020771218 0.020769315 0.020780972
## 
## $future.risk$`5`$Ovarian
##   ByAge estimate lower upper
## 1    70        0     0     0
## 2    75        0     0     0
## 3    80        0     0     0
## 4    85        0     0     0
## 5    90        0     0     0
\end{verbatim}
\end{kframe}
\end{knitrout}

\begin{figure}[h!]
    \centering
    \includegraphics[width=\textwidth]{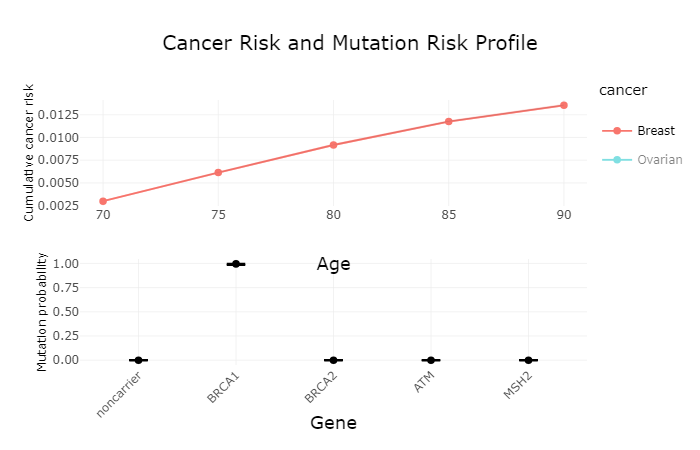}
    \caption{Sample output using \code{visRisk} function.}
    \label{fig:vis_risk}
\end{figure}

The package includes the function \code{visRisk} to visualize the output graphically. Figure~\ref{fig:vis_risk} demonstrates this usage for \code{small.fam}. The \code{visRisk} function was implemented using the \emph{plotly} \citep{plotly} package, so that the output can be rendered interactively and display the exact probabilities upon hovering.

\subsection{Implementation summary}
\label{subsec:implementation}

We list the key functions with their input(s) and output(s) in Table \ref{tab:PanelPRO_functions}. The \code{PanelPRO} function calls the pre-processing functions and the algorithm engine in sequence, so we expect that most users will only need to use this main function. However, the other functions can be called separately if desired.

\begin{table}
\begin{tabularx}{\linewidth}{p{3.5cm}p{3cm}X}
\toprule
Category & Name & Description \\
\midrule 

Pre-processing & \code{buildDatabase} & Subsets the internal database \code{PanelPRODatabase} depending on the cancers and genes selected. The input is the list \code{PanelPRODatabase}. The output is another \code{list} which is a subset of \code{PanelPRODatabase}. \\ [1cm]

Pre-processing & \code{checkFam} & Checks family structure as defined by the user. The inputs are a \code{data.frame} specifying the pedigree and a built database returned by \code{buildDatabase}. The output is a modified \code{data.frame} pedigree and list of imputed ages, if missing ages were imputed (see Section~\ref{subsec:imputation}). \\ [1cm]

Algorithm & \code{PanelPROCalc} & Estimates the posterior carrier probabilities and future risks of the proband. The inputs are the outputs of \code{checkFam}. The outputs are lists of posterior probabilities and future risks for the proband. \\ [1cm]

Main function & \code{PanelPRO} & Runs main function. The inputs are the user-specified pedigree, a vector of cancers in the model, a vector of genes in the model, and other optional parameters. The output is a list of estimates of posterior carrier probabilities for each genotype, along with future cancer risks and ranges for each of these. \\ [1cm]

\bottomrule
\end{tabularx}
\vspace{0.1cm}
\caption{List of functions in \code{PanelPRO}}
\label{tab:PanelPRO_functions}
\end{table}

\section{Discussion}
\label{sec:discussion}

\textbf{PanelPRO} is a highly flexible package which provides an interface to efficiently calculate carrier probabilities for a wide array of cancer susceptibility genes, as well as future cancer risks. It is designed for \Rspace users. Similarly to its ancestors in the \textbf{BayesMendel} package, it can provide the computational engine behind clinical and counseling decision support tools. 

It excels in being fully customizable. Any combination of the 25 genes and 18 cancers currently in the package can be included in the model. New genes and cancers can easily be added, and in fact the code allows for arbitrary number of genes and cancers. Risk modifiers have been included for certain procedures, and more can be added as additional information becomes available. The user can also change the internal database of parameter values. 

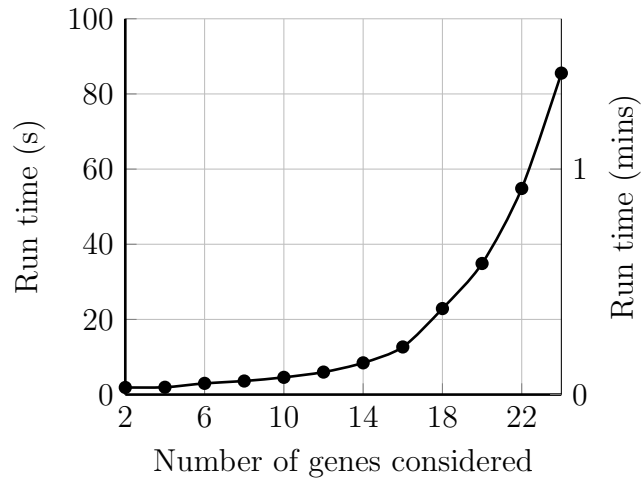
\begin{figure}[h!]
    \centering
    \begin{tikzpicture}
			\begin{axis}[%
			table/col sep=comma,
			height=5cm,
			axis y line*=left,
			scale only axis,
			xmin=2, xmax=24,
			xtick={2, 6, 10, 14, 18, 22},
			xmajorgrids,
			xlabel={Number of genes considered},
			ymin=0.0, ymax=100,
			ylabel={Run time (s)},
			ymajorgrids,
			axis lines*=left,
			line width=1.0pt,
			mark size=2.0pt,
			legend style={at={(0.05,0.95)},anchor=north west,draw=black,fill=white,align=left}]
			
			\addplot [color=black,
			solid,
			mark=*,
			mark options={solid},
			smooth
			] table [x = {nGenes}, y = {PPsec}, col sep = comma] {csvs/PanelPRO_benchmarking.csv};
			
			\end{axis}
			
			\begin{axis}[
    		  height=5cm,
    		  axis y line*=right,
    		  scale only axis,
              axis x line=none,
              ymin=0, ymax=100/60,
              ylabel={Run time (mins)},
              ytick={0, 1},
              xmin=2, xmax=24
            ]
            \end{axis}
			
	\end{tikzpicture}
	\caption{Sample run-times for \code{small.fam} evaluated by \textbf{PanelPRO} on the default settings, as a function of the number of genes considered. The paring parameter is set to 2. These run time experiments were performed on a 2013 Windows machine with an Intel(R) i7-45000 chip at 1.80GHz.}
	\label{fig:run_times}
\end{figure}

The package includes a comprehensive check on the input pedigree to ensure users are informed of potentially inconsistent or infeasible data entries. When it is possible to do so safely, the data is automatically remedied and the user is then notified. Otherwise, the program will halt with an informative error message. Once the pedigree is pre-processed, the posterior probabilities are calculated efficiently. For example, \code{small.fam}, which has 19 members and family history of 2 cancers, runs with all the default settings in a few seconds as shown in Figure~\ref{fig:run_times}. The polynomial run-time of the peeling-paring algorithm is alleviated with \textbf{PanelPRO}'s \textbf{Rcpp} implementation. Even when relaxing the maximum mutations (paring) parameter, $T$, the \Cppspace implementation is able to handle the calculations efficiently. Run-times in these ranges are certainly appropriate for clinical use, as well as use in a research setting where possibly hundreds of pedigrees have to be processed through \textbf{PanelPRO}. Moreover, the peeling-paring algorithm run-time scales linearly in the number of family members in the pedigree and can handle hundreds of members in an inter-generational configuration easily.

 \textbf{PanelPRO} has two main limitations. Firstly, the initial release does not handle pedigrees which contain loops. This additional functionality would be desirable in future releases although loops in pedigrees do not happen frequently. Several studies suggest either exact or approximate computations for pedigrees with loops, see \cite{stricker1995algorithm} and \cite{fernando2009}. Secondly, the polynomial scaling of peeling-paring as a function of the number of genes considered becomes significant when many genes are incorporated. This issue is of concern because we strive for future releases to contain far more genes than 25 as data becomes available. Alternative algorithms which have different time complexity properties, such as the Lander-Green family of algorithms \citep{lander1987construction}, should be explored. These algorithms scale linearly in terms of the number of genes considered, but are exponential in the number of family members in the pedigree \citep{gao2009haplotyping}. A future objective for this package is to contain a choice of the carrier probability calculation method, and ideally an automatic selection of the one which is most efficient, depending on family size and total number of genes. Appropriate thresholds of these two parameters need to be determined by a comprehensive benchmarking exercise.

\section*{Acknowledgements}

We gratefully acknowledge support from the National Cancer Institute at the National Institutes of Health grants 5T32CA009337 (JL and TH), 2T32CA009001 (TH), and 4P30CA006516 (GP). 

\clearpage

\bibliographystyle{abbrvnat}
\bibliography{bibliography}



\end{document}